\documentclass[journal]{IEEEtran}

\usepackage[cmex10]{amsmath}
\usepackage[utf8x]{inputenc}
\usepackage[T1]{fontenc}
\usepackage[american]{babel}
\usepackage[keeplastbox]{flushend}
\usepackage{amssymb, bm, graphicx, ucs, cite, booktabs, algorithmic, algorithm, color}

\IEEEoverridecommandlockouts

\title{Spatial Diversity in Radar Detection via \\ Active Reconfigurable Intelligent Surfaces}

\author{Mohamed~Rihan,~\IEEEmembership{Senior~Member,~IEEE}, Emanuele~Grossi,~\IEEEmembership{Senior~Member,~IEEE}, \\
Luca~Venturino,~\IEEEmembership{Senior Member,~IEEE}, Stefano~Buzzi,~\IEEEmembership{Senior Member,~IEEE}, 
 \thanks{The work of E. Grossi and L. Venturino was supported by the research program ``Dipartimenti di Eccellenza 2018--2022'' sponsored by the Italian Ministry of Education, University, and Research (MIUR). The work of M. Rihan and S. Buzzi was supported by H2020 Marie Sk\l{}odowska-Curie Actions (MSCA) Individual Fellowships (IF) RASECOL, grant agreement 898354.}
 \thanks{The authors are with the Department of Electrical and Information Engineering (DIEI), University of Cassino and Southern Lazio, 03043 Cassino, Italy (e-mails: mohamed.elmelegy@el-eng.menofia.edu.eg, e.grossi@unicas.it, l.venturino@unicas.it, buzzi@unicas.it).}
 \thanks{M. Rihan is on leave from the Faculty of Electronic Engineering, Menoufia University, Egypt. E. Grossi, L. Venturino, and S. Buzzi are also with Consorzio Nazionale Interuniversitario per le Telecomunicazioni, 43124 Parma, Italy (e-mail: e.grossi@unicas.it, l.venturino@unicas.it, buzzi@unicas.it).}
}

\begin{document}

\bstctlcite{BSTcontrol}
\maketitle

\begin{abstract}
Active reconfigurable intelligent surfaces (RISs) are a novel and promising technology that allows controlling the radio propagation environment while compensating for the product path loss along the RIS-assisted path. In this letter, we consider the classical radar detection problem and propose to use an active RIS to get a second independent look at a prospective target illuminated by the radar transmitter. At the design stage, we select the power emitted by the radar, the number of RIS elements, and their amplification factor in order to maximize the detection probability for a fixed probability of false alarm and a common (among radar and RIS) power budget. An illustrative example is provided to assess the achievable detection performance, also in comparison with that of a radar operating alone or with the help of a passive RIS.
\end{abstract}


\section{Introduction}

\IEEEPARstart{R}{econfigurable} intelligent surfaces (RISs) allow controlling the radio propagation environment, thus providing novel degrees of freedom for the design of wireless systems~\cite{liu2021reconfigurable, Prasad2021}. An RIS is a low-cost passive flat surface made of sub-wavelength refractive/reflective elements (atoms) that can add a tunable phase shift to the incident electromagnetic wave. The atoms are controlled using an embedded logic with a power consumption that is usually negligible and, all together, can redirect the planar or spherical wavefront hitting the surface in several ways. E.g., a diffuse scattering, an anomalous reflection, or a beam focused towards a specific point can be produced; in addition, a data message can be even superimposed on the redirected signal~\cite{Guo-2020, Dai-2021}. RISs have been used to boost the performance of wireless communication links~\cite{Geoffrey-Ye-2020, Pan-2021, Basar-2021}; they have been also proven effective in other contexts, including wireless power transfer~\cite{zhao2020wireless}, localization and mapping~\cite{2020-Wymeersch-Localization-and-Mapping, Alouini-Localization}, joint communication and sensing~\cite{joint_waveform}, and, more recently, target detection~\cite{Grossi2021ris, Aubry-2021, foundations}.

A major drawback of passive RISs is that the end-to-end indirect link (source $\rightarrow$ RIS $\rightarrow$ destination) presents a heavy product path-loss attenuation (also referred to as double fading attenuation), which may limit their usability, especially if a direct link is available~\cite{Najafi-2021}. In particular, recent studies suggest that such passive devices should be better placed close to the transmitter or the receiver~\cite{Dunna-2020, Basar-2021, foundations}. To improve the end-to-end power budget of an RIS-assisted communication system, the works in~\cite{Larsson-2021, ActiveRIS2021} have recently introduced the idea of using an \emph{active} RIS, wherein each reflecting element employs an active load to amplify the incident signal. A shortcoming is that the circuitry controlling such an active RIS introduces an internal source of noise and consumes a non-negligible amount of power; however, even after accounting for these effects in the overall link budget, the results in~\cite{Larsson-2021, ActiveRIS2021} indicate that the communication system can perform much better if aided by an active rather than a passive RIS.

\begin{figure}[t]
 \centerline{\includegraphics[width=\columnwidth]{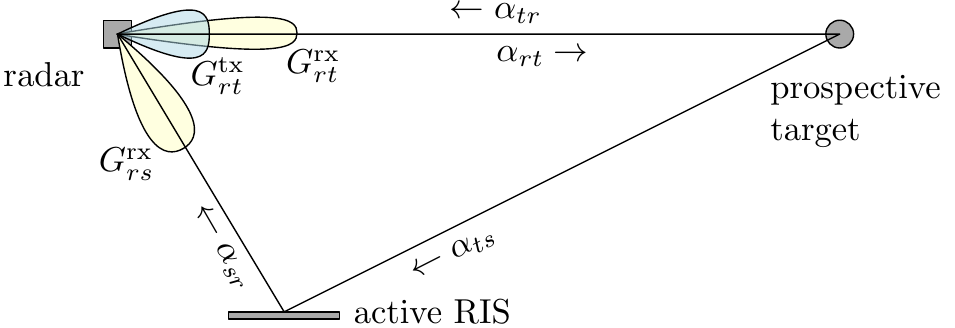}}
 \vspace{-5pt}\caption{Considered architecture composed of a radar, equipped with one transmit and two receive beams, aided by an active RIS; radar and RIS are widely-spaced with respect to the target.}
 \label{fig_1}
\end{figure} 
In this work, we propose to use an active RIS to improve the detection capability of a radar system. The main intuition is that an active RIS can offer a second look at a target illuminated by the radar, thus providing spatial (angular) diversity, \emph{and}, in addition, can compensate for the product path loss along the indirect target-RIS-radar path. Overall, the proposed RIS-assisted architecture may realize a sort of low-cost distributed radar system, as illustrated in Fig.~\ref{fig_1}; in particular, rather than having a second radar receiver equipped with a full radio-frequency processing chain and using a dedicated data link from each receiver to a common fusion center, an active RIS simply redirects the impinging signal towards a \emph{unique} destination, which collects both the direct and indirect target echoes through two dedicated spatial beams. More specifically, we make the following contributions. Upon adapting the signal model developed in~\cite{Larsson-2021, ActiveRIS2021} to the considered application, we propose to choose the number of RIS elements, their amplification gain, and the power split among the radar transmitter and the active RIS to maximize the detection probability for a fixed probability of false alarm. Since this problem is non-convex, we derive a suboptimal solution based upon an alternating maximization. The results show that the use of an active RIS can grant a large performance improvement compared to the cases where no RIS or a passive RIS is used.

The remainder of this work is organized as follows. In the next section, the system model is presented, while, in Sec.~\ref{sys_des_sec}, the system design is described. An illustrative example is provided in Sec.~\ref{num_res_sec}, and concluding remarks are given in Sec.~\ref{concl_sec}.

\section{System Model}

Consider a target detection problem, where the radar is assisted by an active RIS, as shown in Fig.~\ref{fig_1}. The radar emits an average power $P_r$ and is equipped with one transmit beam with gain $G^\text{tx}_{rt}$ pointing towards the prospective target and two receive beams, one with gain $G^\text{rx}_{rt}$ pointing towards the target and one with gain $G^\text{rx}_{rs}$ pointing towards the RIS. The radar radiates a passband waveform with bandwidth $W$, duration $T$, and carrier wavelength $\lambda$. The active RIS is composed of $L$ elements and is \emph{widely-spaced} from the radar, so as to get an independent look at the target; in particular, radar and RIS are in each other's far-field. The (complex) amplitude response of the $\ell$-th RIS element is denoted by $a_\ell \mathrm e^{\mathrm i\phi_\ell}$, where $a_\ell \in[1, a_\text{max}]$, with $a_\text{max}\geq 1$, and $\phi_\ell\in [0,2 \pi]$. 
We assume that the target is in the far-field of both the radar and the RIS. We denote by $d_{rt}$, $d_{ts}$, and $d_{sr}$ the radar-target, target-RIS, and RIS-radar distances, respectively; also, we denote by $G_{st}$ and $G_{sr}$ the gain of each element of the RIS towards the target and the radar, respectively. Accordingly, we can define the following positive quantities
\begin{align}
\alpha_{rt} & = \sqrt{G^\text{tx}_{rt}/(4\pi)} / d_{rt}, & \alpha_{tr} &= \lambda \sqrt{G^\text{rx}_{rt}} /(4\pi d_{rt})\\
\alpha_{ts} & = \lambda \sqrt{G_{st}}/(4\pi d_{ts}), & \alpha_{sr} & = \lambda \sqrt{G_{sr} G^\text{rx}_{rs}} /(4\pi d_{sr})
\end{align}
which account for the link budget along each hop in Fig.~\ref{fig_1}.

After base-band conversion, matched-filtering with the transmit waveform, and sampling,\footnote{The signal from the beam pointing towards the target is sampled at the delay $2d_{rt}/c$, where $c$ is the speed of light; similarly, the signal from the beam pointing towards the RIS is sampled at the delay $(d_{rt}+d_{ts}+d_{sr})/c$.} if a target is present in the cell under test, the baseband signal received by the radar beam pointing toward the target can be written as
\begin{equation}
y_1 = \gamma_1 \alpha_1\sqrt{P_r} + w_1
\end{equation}
where $\gamma_1\in\mathbb C$ is the unknown target response towards the radar, $\alpha_1 = \alpha_{rt} \alpha_{tr} \sqrt{WT}$, and $w_1 \sim \mathcal{CN}(0, P_{w_1})$ is the noise, distributed as a complex circularly symmetric Gaussian random variable with variance $P_{w_1}$. Also, upon denoting by $\beta_{t\ell}$ and $\beta_{\ell r}$ the phase delays along the path linking the target and the $\ell$-th RIS element and the path linking the $\ell$-th RIS element and the radar, respectively, the baseband signal received by the radar beam pointing towards the RIS is
\begin{align}
y_2 & = \gamma_2 \alpha_2 \sqrt{P_r} \sum_{\ell=1}^L a_\ell \mathrm e^{\mathrm i(\beta_{t\ell}+\phi_\ell+\beta_{\ell r})}\notag\\
&\quad + \alpha_{sr} \sum_{\ell=1}^L a_\ell v_\ell \mathrm e^{\mathrm i(\phi_\ell+\beta_{\ell r})} + w_2 \label{eq_y2}
\end{align}
where $\gamma_2 \in \mathbb C$ is the unknown target response towards the RIS, $\alpha_2 = \alpha_{rt} \alpha_{ts} \alpha_{sr} \sqrt{WT}$, $v_\ell\sim \mathcal{CN}(0,P_v)$ is the \emph{dynamic noise} generated by the $\ell$-th RIS element~\cite{Larsson-2021, ActiveRIS2021}, and $w_2 \sim \mathcal{CN}(0, P_{w_2})$, independent of $w_1$, is the receive noise. 

We assume that $\{v_\ell\}_{\ell=1}^L$ are independent, so that the distribution of $\sum_{\ell=1}^L a_\ell v_\ell \mathrm e^{\mathrm i(\phi_\ell+\beta_{\ell r})}$ is not influenced by $\{\phi_\ell\}_{\ell=1}^L$. Also, since $\beta_{t\ell}$ and $\beta_{\ell r}$ are known,\footnote{($\{\beta_{t\ell}\}_{\ell=1}^L$ and $\{\beta_{\ell r}\}_{\ell=1}^L$ are uniquely determined by the orientation of the RIS and by the mutual position of radar, RIS, and target; see also~\cite{Grossi2021ris}.} the RIS phases can be chosen as $\phi_\ell = -\beta_{t\ell} - \beta_{\ell r}$, so that all the signal terms in~\eqref{eq_y2} are \emph{phase aligned} (see also~\cite{Grossi2021ris}). In this case, Eq.~\eqref{eq_y2} becomes
\begin{equation}
y_2 = \gamma_2 \alpha_2 \sqrt{P_r} \sum_{\ell=1}^L a_\ell + z_2 \label{y_2_eq}
\end{equation}
where $z_2= w_2 + \alpha_{sr} \sum_{\ell=1}^L a_\ell v_\ell \mathrm e^{-i\beta_{t\ell}}$ is distributed as $\mathcal{CN}(0, P_{w_2}+ \alpha^2_{sr}P_v\sum_{\ell=1}^L a_\ell^2)$.

Following~\cite{Larsson-2021, ActiveRIS2021}, we model the RIS power consumption as $P_s = L\rho_s + \eta_s^{-1} p_\text{out}$, where $\rho_s=P_c +P_{dc}$ is the power consumption needed to operate each reflecting element of the RIS, with $P_{c}$ the switch and control circuit power and $P_{dc}$ the DC biasing power consumption, $\eta_s$ is the amplifier efficiency, and $p_\text{out}$ is the output power. This latter term is in turn given by $(\alpha^2_{rts} \sigma^2_{\gamma_2} P_r + P_v)\sum_{\ell=1}^L a_\ell^2$, where $\sigma^2_{\gamma_2}$ is the mean square value of $\gamma_2$ and $\alpha_{rts}= \alpha_{rt} \alpha_{ts}$, so that
\begin{equation}
P_s = L\rho_s + \eta_s^{-1} (\alpha^2_{rts} \sigma^2_{\gamma_2} P_r + P_v)\sum_{\ell=1}^L a_\ell^2.
\end{equation}

\section{System design}\label{sys_des_sec}

It is not difficult to show that the generalized likelihood ratio test~\cite{Van_Trees_1} with respect to the unknown target responses $(\gamma_1, \gamma_2)$ based on the observations $(y_1, y_2)$ is
\begin{equation}
\frac{|y_1|^2}{P_{w_1}} + \frac{|y_2|^2}{P_{w_2}+ \alpha^2_{sr}P_v\sum_{\ell=1}^L a_\ell^2} \gtrless \gamma 
\label{GLRT} 
\end{equation}
and a target is declared if the detection threshold $\gamma$ is exceeded. The probability of false alarm is $\text{PFA}=(1+\gamma) \mathrm e^{-\gamma}$~\cite{Richards_2005}, and $\gamma$ is usually set to have a specified PFA level. As to the probability of detection (PD), we need to specify the joint distribution of the target responses: if $\gamma_1\sim\mathcal{CN}(0,\sigma^2_{\gamma_1})$ and $\gamma_2\sim\mathcal{CN}(0,\sigma^2_{\gamma_2})$ are independent, we have\footnote{This corresponds to the Swerling's Case~1, and the test statistic in~\eqref{GLRT} is the sum of two independent exponential random variables with mean $1+\text{SNR}_1$ and $1+\text{SNR}_2$, that follows a hypo-exponential density~\cite{Ross_2014}.}
\begin{equation}
\text{PD}= \frac{1+\text{SNR}_1}{\text{SNR}_1-\text{SNR}_2} \mathrm e^{-\frac{\gamma}{1+\text{SNR}_1}} - \frac{1+\text{SNR}_2}{\text{SNR}_1-\text{SNR}_2} \mathrm e^{-\frac{\gamma}{1+\text{SNR}_2}} \label{Pd}
\end{equation}
where
% \begin{subequations}\label{SNRs}
% \begin{align}
% \text{SNR}_1 & = \frac{\alpha_1^2 \sigma^2_{\gamma_1} P_r}{P_{w_1}} \label{SNR_1}\\
% \text{SNR}_2 & = \frac{\alpha_2^2 \sigma^2_{\gamma_2} P_r \bigl(\sum_{\ell=1}^L a_\ell\bigr)^2}{P_{w_2}+ \alpha^2_{sr}P_v\sum_{\ell=1}^L a_\ell^2} \label{SNR_2}
% \end{align}%
% \end{subequations}
\begin{equation}
 \text{SNR}_1 = \frac{\alpha_1^2 \sigma^2_{\gamma_1} P_r}{P_{w_1}}, \quad 
 \text{SNR}_2 = \frac{\alpha_2^2 \sigma^2_{\gamma_2} P_r \bigl(\sum_{\ell=1}^L a_\ell\bigr)^2}{P_{w_2}+ \alpha^2_{sr}P_v\sum_{\ell=1}^L a_\ell^2}\label{SNRs}
\end{equation}
are the SNRs on the two receive channels.

Our goal here is to maximize PD for a fixed PFA. The available degrees of freedom for system optimization are the radar power $P_r$, the number $L$ of RIS elements, and the corresponding amplification factors $\bm a = (a_1 \cdots a_L)$; instead, the physical constraints are on the maximum number of RIS elements $L_\text{max}$, on the maximum amplification factor $a_\text{max}$, and on the overall power budget $\rho_r + \eta_r^{-1}P_r + P_s$, where $\rho_r$ is the circuit power required to operate the radar transmitter, and $\eta_r$ is the efficiency of the radar amplifier. Hence, the optimization problem tackled here is
\begin{equation} \label{opt_prob}
\begin{aligned}
\max_{L\in\mathbb N} \max_{\bm a \in \mathbb R^L, P_r \in\mathbb R} & \; \text{PD}\bigl(\text{SNR}_1(P_r), \text{SNR}_2(L, \bm a, P_r) \bigr)\\
\text{s.t.} & \; \rho_r +\eta_r^{-1} P_r + L\rho_s \\
&\; + \eta_s^{-1} ( \alpha^2_{rts} \sigma^2_{\gamma_2} P_r + P_v) \sum_{\ell=1}^L a_\ell^2 \leq P_\text{max} \\
& \; 1 \leq a_\ell \leq a_\text{max}, \quad \ell=1,\ldots, L \\
& \; 0 \leq L \leq L_\text{max}
\end{aligned}
\end{equation}
where the dependency of the objective function from the optimization variables has been made explicit, while $L=0$ simply means here that the RIS is not used.

The optimization problem in~\eqref{opt_prob} is quite difficult, and we resort to a block-coordinate ascent method, also known as alternating-maximization, where, at each iteration, the objective function is maximized over a \emph{block} of variables, while keeping the others fixed at their previous values. In particular, we consider two reduced-complexity sub-problems: the maximization over $P_r$ and the maximization over the pair $(\bm a,L)$. In the following, these two sub-problems are optimally solved, and a closed-form expression for their solutions is provided in Eqs.~\eqref{P_r_opt} and~\eqref{a_L_opt}. For the reader's sake, the complete routine is reported in Alg.~\ref{alg}.

\subsection{Maximization over the radar power}

Since the probability of detection in~\eqref{Pd} is increasing with the SNRs in~\eqref{SNRs}, that are in turn increasing with the radar transmit power, we should choose the largest value of $P_r$ satisfying the power constraint in~\eqref{opt_prob}, i.e., 
\begin{align}
P_r^{\star} = \frac{P_\text{max} - \rho_r -L\rho_s - \eta_s^{-1}P_v \sum_{\ell=1}^L a_\ell^2}{\eta_r^{-1}+ \eta_s^{-1} \alpha_{rts}^2 \sigma_{\gamma_2}^2 \sum_{\ell=1}^L a_\ell^2}. \label{P_r_opt}
\end{align}

\subsection{Maximization over the RIS parameters}

Since the probability of detection in~\eqref{Pd} is increasing with the SNRs in~\eqref{SNRs}, and since $\text{SNR}_1$ is independent of $(\bm a, L)$, the problem to be solved here is equivalent to
\begin{equation} \label{subprob_L_a}
\begin{aligned}
\max_{L\in\mathbb N} \max_{\bm a \in \mathbb R^L} & \; \text{SNR}_2(L, \bm a, P_r)\\
\text{s.t.} & \; \rho_s L + \zeta \sum_{\ell=1}^L a_\ell^2 \leq \kappa \\
& \; 1 \leq a_\ell \leq a_\text{max}, \quad \ell=1,\ldots, L \\
& \; 0 \leq L \leq L_\text{max}
\end{aligned}
\end{equation}
where $\kappa= P_\text{max}-\rho_r-\eta_r^{-1}P_r$, and $\zeta = \eta_s^{-1} ( \alpha^2_{rts} \sigma^2_{\gamma_2} P_r + P_v)$. We point out here that a similar problem has been tackled in~\cite{Larsson-2021}; however, differently from~\cite{Larsson-2021}, we are now considering additional constraints on $L$ and $a_\ell$, that will result into a slightly different solution, as briefly outlined next. 

Notice first that, the constraints in~\eqref{subprob_L_a} imply that we must necessarily have $L\leq \bar L$, where 
\begin{equation}
\bar L = \min \left\{ L_\text{max}, \left\lfloor \frac{\kappa}{\rho_s+\zeta} \right\rfloor \right\}.
\end{equation}
At this point, from the expression of $\text{SNR}_2$ in~\eqref{SNRs}, it is not difficult to show that, for any $L\in\bigl\{0, 1, \ldots, \bar L \bigr\}$, the maximization over $\bm a$ gives $a_1 = \cdots = a_L = g(L)$, where 
\begin{equation}
g(L) = \begin{cases}
\min\left\{ a_\text{max}, \sqrt{\frac{\kappa - \rho_s L}{\zeta L}} \right\}, & \text{if } L\geq 1\\
1, & \text{if } L=0.
\end{cases}\label{eq_a_L}
\end{equation}
Therefore, Problem~\eqref{subprob_L_a} reduces to
\begin{equation} \label{subprob_L}
\max_{L\in\mathbb N :\; 0 \leq L \leq \bar L} \;\;\frac{\alpha_2^2 \sigma^2_{\gamma_2} P_r L^2 g^2(L) }{P_{w_2}+ \alpha^2_{sr}P_v L g^2(L)} .
\end{equation}

\begin{algorithm}[t]
 \caption{Block-coordinate descent for Problem~\eqref{opt_prob} \label{alg}}
 \begin{algorithmic}
 \STATE choose a feasible triplet $(P_r, \bm a, L)$
 \REPEAT
 \STATE update $P_r$ with~\eqref{P_r_opt}
 \STATE update $(\bm a, L)$ with~\eqref{a_L_opt}
 \UNTIL convergence
 \end{algorithmic}
\end{algorithm}
Let us now relax Problem~\eqref{subprob_L} by extending the search set to $\bigl\{L\in \mathbb R : 0 \leq L \leq \bar L\bigr\}$. Let $f(L)$ and $f'(L)$ denote the objective function of~\eqref{subprob_L} and its first order derivative, respectively, and define $L_1 = \kappa/(\rho_s + a^2_\text{max}\zeta)$. 
Then, from~\eqref{eq_a_L}, we have that, if $0\leq L\leq L_1$,
\begin{equation}
f(L) = \frac{\alpha_2^2 \sigma^2_{\gamma_2} P_r a^2_\text{max}L^2}{P_{w_2}+ \alpha^2_{sr}P_v a^2_\text{max}L}
\end{equation}
and $f'(L)\geq 0$; if, instead, $L\geq L_1$,
\begin{equation}
f(L) = \frac{\alpha_2^2 \sigma^2_{\gamma_2} P_r L (\kappa - \rho_s L)}{P_{w_2} \zeta + \alpha^2_{sr} P_v (\kappa - \rho_s L)}
\end{equation}
and $f'(L)\geq 0$ for $L\leq L_2^-$ or $L\geq L_2^+$, and $f'(L) \leq 0$ for $L_2^-\leq L \leq L_2^+$, where
\begin{equation}
 L_2^\pm =\frac{P_{w_2} \zeta + \alpha^2_{sr} P_v \kappa \pm \sqrt{\bigl(P_{w_2} \zeta+ \alpha^2_{sr} P_v \kappa\bigr) P_{w_2}\zeta}}{\alpha^2_{sr} P_v \rho_s}
\end{equation}

Notice now that $\max \bigl\{ L_2^-, \bar L\bigr\} \leq \kappa/\rho_s \leq L_2^+$. Therefore, if $L_1\leq L_2^-$, we have $f'(L)\geq 0$ for $L\leq L_2^-$, and $f'(L)\leq 0$ for $L_2^-\leq L \leq \kappa/\rho_s$; then, the solution to the relaxed problem is $L^\star_\text{rel}= \min \bigl\{ L_2^-, \bar L \bigr\}$. If, instead, $L_1\geq L_2^-$, we have $f'(L)\geq 0$ for $L\leq L_1$, and $f'(L)\leq 0$ for $L_1\leq L \leq \kappa/\rho_s$, so that the solution to the relaxed problem is $L^\star_\text{rel}= \min \bigl\{L_1, \bar L \bigr\}$. Thus
\begin{equation}
 L^\star_\text{rel} = \min \bigl\{ \max \{ L_1, L_2^-\}, \bar L \bigr\}
\end{equation}

\begin{figure*}[t]
 \centering
 \centerline{\includegraphics[width=0.33\textwidth]{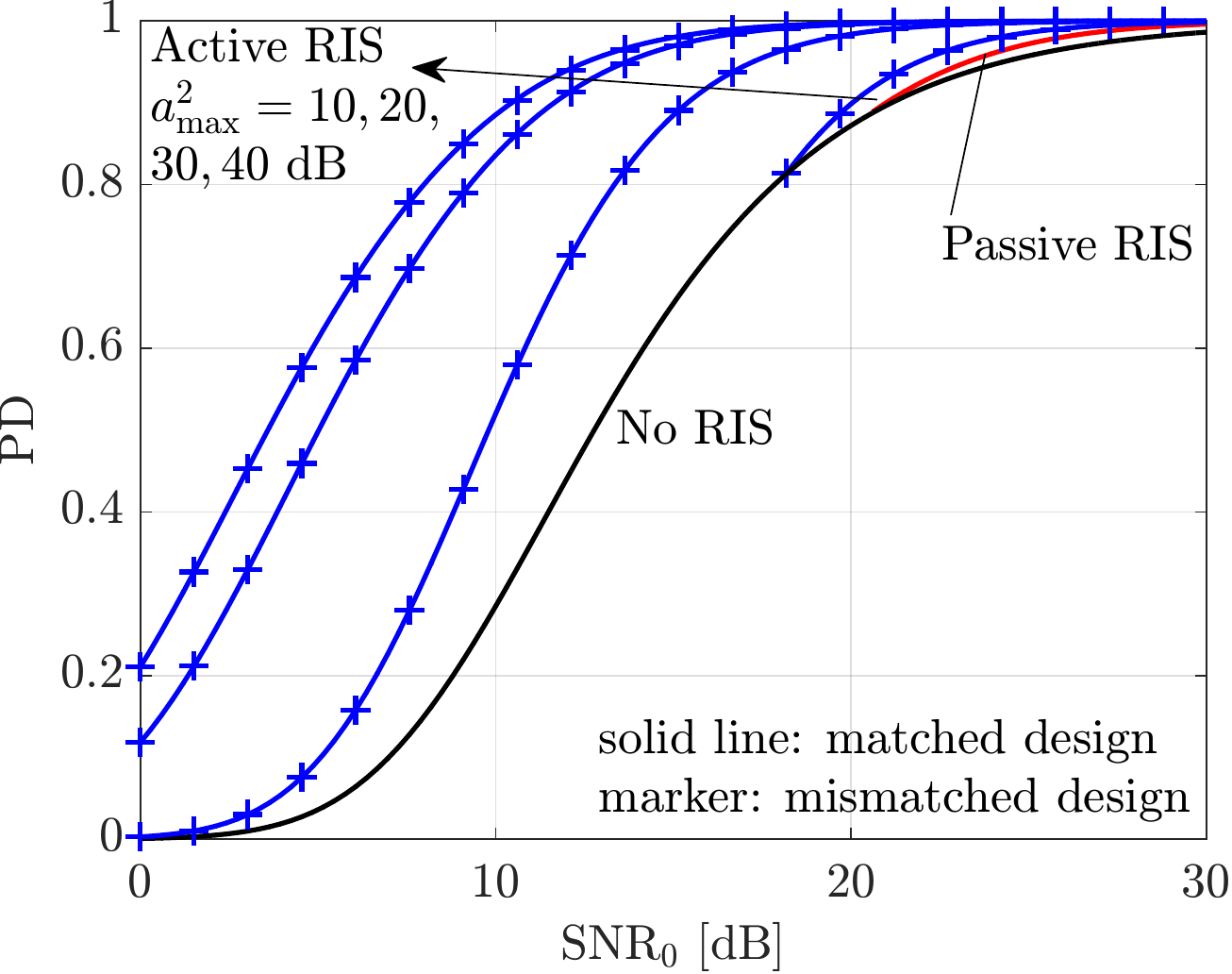}
 \includegraphics[width=0.33\textwidth]{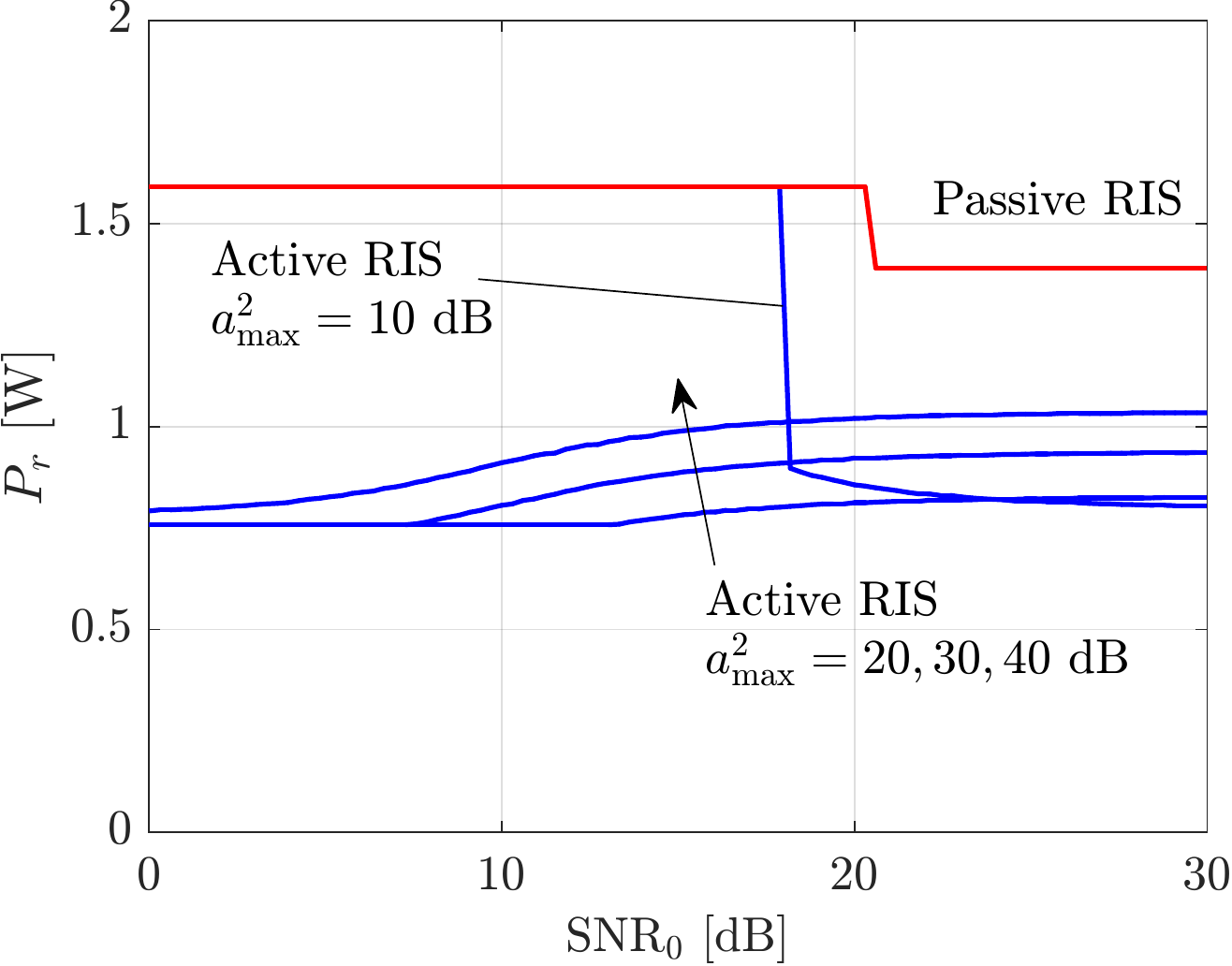}
 \includegraphics[width=0.33\textwidth]{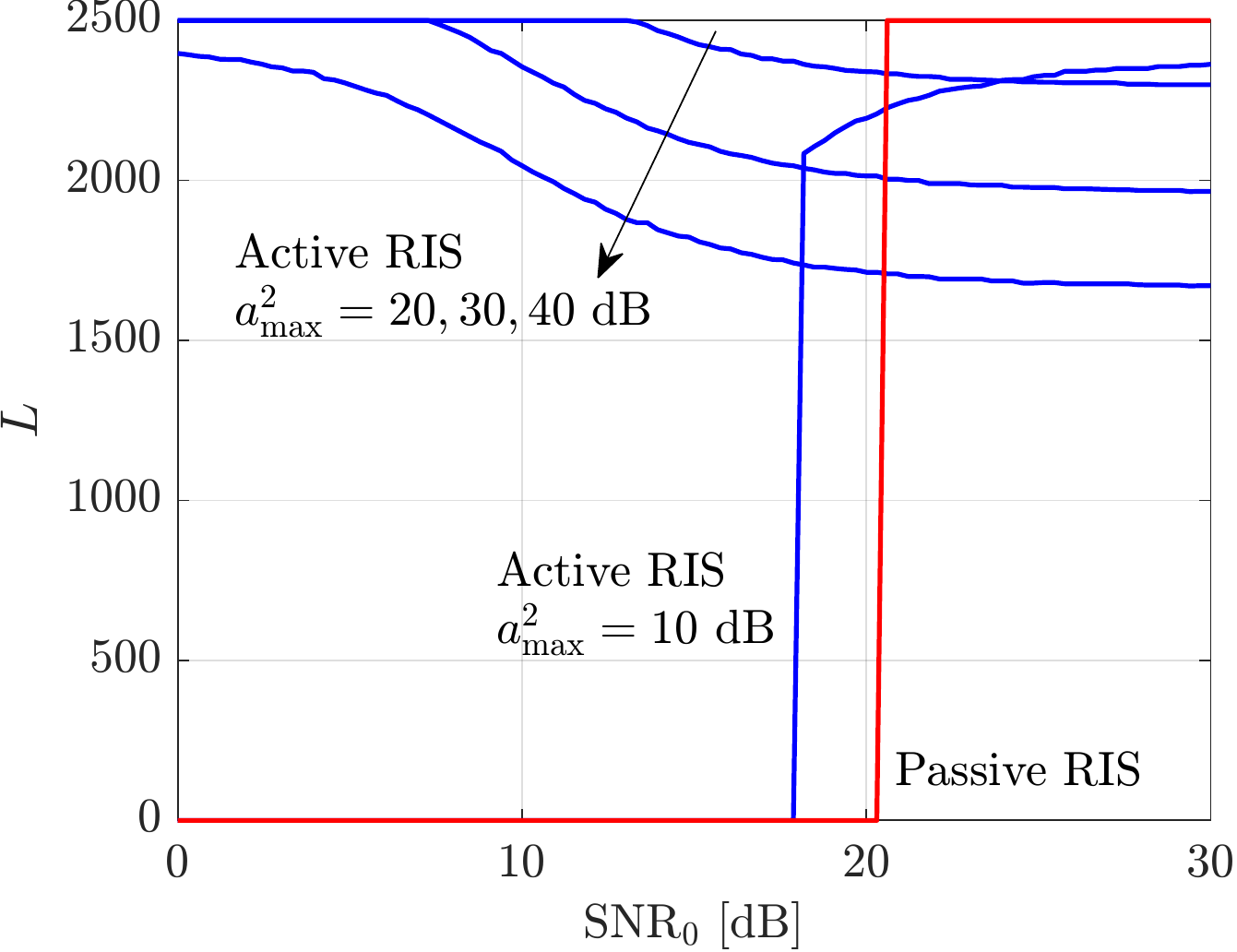}}
 \caption{Probability of detection (left plot), radar transmit power (middle plot), and RIS element number (right plot) versus $\text{SNR}_0$ for a radar operating alone (``No RIS''), aided by a passive RIS, or helped by an active RIS with a maximum amplification factor of $a^2_\text{max}=10,20,30,40$ dB.\label{fig_2}}
\end{figure*}

Finally, the solution to Problem~\eqref{subprob_L} is found by comparing $f\bigl(\lfloor L_\text{rel}^\star \rfloor \bigr)$ and $f\bigl(\lceil L_\text{rel}^\star \rceil\bigr)$; consequently, the solution to Problem~\eqref{subprob_L_a} is
\begin{subequations} \label{a_L_opt}
 \begin{align}
 L^\star &= \operatorname*{argmax}\limits_{L\in\{ \lfloor L_\text{rel}^\star \rfloor, \lceil L_\text{rel}^\star \rceil\}} \frac{\alpha_2^2 \sigma^2_{\gamma_2} P_r L^2 g^2(L) }{P_{w_2}+ \alpha^2_{sr}P_v L g^2(L)}\\
 a_\ell^\star &= g(L^\star), \quad \ell=1,\ldots, L^\star.
 \end{align}%
\end{subequations}

\section{Simulation Results} \label{num_res_sec}

Here we consider an illustrative example to assess the performance improvement granted by an active RIS in the radar detection problem. With reference to Fig.~\ref{fig_1}, the radar is located at the origin of the coordinate system, while the target and the RIS lie in positions $(500,0)$ m and $(200,-200)$ m, respectively; the system parameters are listed in Table~\ref{tab_1}. For comparison, we also consider the cases where no RIS is used (therefore, $P_r=(P_\text{max}-\rho_r)\eta_r$) and where a passive RIS is adopted (therefore, $\rho_s=P_c$, $P_v=0$, $\eta_s=1$, and $a_\ell=1$ for any $\ell$) and optimized as in Problem~\eqref{opt_prob}, with power constraint $\rho_r +\eta_r^{-1} P_r + LP_c \leq P_\text{max}$. All the curves are reported versus $\text{SNR}_0=\alpha_1^2 \sigma^2_{\gamma_1} (P_\text{max}-\rho_r)\eta_r/P_{w_1}$ (i.e., the SNR in the No RIS case), where the target strength $\sigma^2_{\gamma_1}$ is varied, with $\sigma^2_{\gamma_2}=\sigma^2_{\gamma_1}$.

\begin{table}[t]
 \centering 
 \caption{System parameters \label{tab_1}}
 \begin{tabular}{lll}
 \toprule
 $P_\text{max}= 4$ W & $W=10$ MHz & $\eta_r = \eta_s = 0.8$ \\
 $a^2_\text{max}\in\{10, 20, 30, 40\}$ dB & $T=0.5$ ms & $\rho_r=2$ W\\
 $G_{st}= G_{sr} =3$ dB & $\lambda = 10$ cm & $P_c = -10$ dBm\\
 $P_{w_1}= P_{w_2}= -128$ dBm & $\text{PFA} = 10^{-6}$ & $P_{dc}=- 5$ dBm \\
 $G^\text{tx}_{rt} = G^\text{rx}_{rt} = G^\text{rx}_{rs} = 33$ dB & $L_\text{max} =2500$ & $P_v= -134$ dBm\\
 \bottomrule
 \end{tabular}
\end{table}

In the left plot of Fig.~\ref{fig_2}, the optimized PD is reported (solid lines). It is seen by inspection that the use of a passive RIS is advantageous (compared to the No RIS case) only for very large SNRs (corresponding in this example to PD values larger than 0.89), in accordance with the findings in~\cite{Grossi2021ris}. The active RIS with a maximum amplification factor of 10 dB only performs slightly better than the passive RIS and is still not competitive (compared to the No RIS case) over a large SNR range; on the contrary, the active RIS with a maximum amplification factor greater than or equal to 20 dB outperforms the passive RIS and the No RIS cases in all the inspected range of SNRs; remarkably, if $a^2_\text{max}=40$ dB, a gain of as much as 9.2~dB at $\text{PD}=0.5$ can be achieved compared to the No RIS case. Notice that the proposed solution requires prior knowledge of $\sigma^2_{\gamma_2}$, which is not available in practice. A viable fix is to use a design value, say $\sigma^2_{\gamma_2,d}$, and accept some loss in case of mismatch. In this figure, we have also reported PD for such mismatched design ($+$ marker), where $\sigma^2_{\gamma_2,d}$ is the one corresponding to $\text{PD}=0.5$. As it can be seen, a negligible loss is incurred for $a^2_\text{max}= 20, 30, 40$ dB, showing that this design is robust as to the uncertainty in the target strength.

In the middle and right plots of Fig.~\ref{fig_2}, instead, the optimal radar transmit power and number of RIS elements are shown, respectively. The optimal amplification factor is not reported since, for the considered system, $a_\ell = a^2_\text{max}$ for all $\ell$ and in all cases. It is seen that the passive RIS is activated only for $\text{SNR}_0\geq 20.5$ dB: in this latter case, all reflecting elements are used, with a power consumption of about $L_\text{max}P_c=0.25$ W. The active RIS with a maximum amplification factor of 10 dB behaves similarly to the passive RIS case: indeed here the signal amplification is still not sufficient to cope with the severe product path loss along the indirect target-RIS-radar hop. If instead the maximum amplification factor is set to 20 dB, the radar power consumption is $\rho_r+\eta_r^{-1}P_r\approx3$~W; interestingly, as $\text{SNR}_0$ increases, the power is slightly moved here from the RIS to the radar side, as the number of required RIS elements decreases. When $a^2_\text{max}$ is further increased, the number of RIS elements is progressively reduced for the same value of $\text{SNR}_0$ and, consequently, the power consumption of the RIS decreases: the intuition here is that, once the RIS amplification is already sufficiently large to well counteract the product path loss along the target-RIS-radar path, it becomes more advantageous to switch off more RIS elements and move the power at the radar transmitter to better illuminate the target.

\section{Conclusion}\label{concl_sec}
In this letter, we have presented a novel radar architecture, where the radar transmitter illuminates a prospective target, and the radar receiver collects the \emph{direct} echo and an \emph{indirect} echo bouncing on an active RIS widely-spaced from the radar. This system allows to exploit the spatial (angular) diversity of the target response, and, once the RIS is properly sized and the available power properly split among the radar transmitter for target illumination and the RIS for path loss compensation, a significant improvement in the detection probability can be obtained compared to the case where the radar operates alone or with the help of a passive RIS. Future developments may include the use of multiple RISs, the extension to the case where the noise and/or clutter power is unknown and must be estimated (through, e.g., adaptive techniques~\cite{Guerci_2014, Liu_2022}), and the study of joint detection and estimation procedures.

% Generated by IEEEtran.bst, version: 1.14 (2015/08/26)

\end{document}